\documentclass[sigconf,natbib=true]{acmart}

\setcopyright{none}
\renewcommand\footnotetextcopyrightpermission[1]{}  

\AtBeginDocument{%
  }

\acmConference[LLM4Eval '25]{Third Workshop on Large Language Models for Evaluation in Information Retrieval (LLM4Eval 2025) co-located with the 48th International Conference on Research and Development in Information Retrieval (SIGIR 2025)}{July 17, 2025}{Padua, Italy}
\acmISBN{}

\usepackage{multirow}
\usepackage{tcolorbox}
\tcbuselibrary{breakable}

\newtcolorbox[auto counter, number within=section]{promptbox}[2][]{
    colback=gray!00, 
    colframe=black, 
    width=\textwidth, 
    arc=1.5mm, auto outer arc, 
    breakable, 
    left=0.9mm, right=0.9mm, 
    boxrule=0.9pt, 
    title={\thetcbcounter: #1}, 
    label={#2}, 
}

\newcommand{\Autoref}[1]{%
  \begingroup%
  \def\sectionautorefname{Section}%
  \def\subsectionautorefname{Section}%
  \def\subsubsectionautorefname{Section}%
  \autoref{#1}%
  \endgroup%
}
\begin{document}

\title[Medical Case-based Judgments with Multimodal LLMs]{Expanding Relevance Judgments for Medical Case-based~Retrieval Task with Multimodal LLMs}


\author{Catarina Pires}
\orcid{0009-0005-5000-6333}
\email{up201907925@fe.up.pt}
\affiliation{%
  \institution{INESC TEC and Faculty of Engineering, University of Porto}
  \city{Porto}
  \country{Portugal}
}

\author{Sérgio Nunes}
\orcid{0000-0002-2693-988X}
\email{ssn@fe.up.pt}
\affiliation{%
  \institution{INESC TEC and Faculty of Engineering, University of Porto}
  \city{Porto}
  \country{Portugal}
}

\author{Luís Filipe Teixeira}
\orcid{0000-0002-4050-7880}
\email{luisft@fe.up.pt}
\affiliation{%
  \institution{INESC TEC and Faculty of Engineering, University of Porto}
  \city{Porto}
  \country{Portugal}
}

\renewcommand{\shortauthors}{Pires et al.}

\begin{abstract}
    Evaluating Information Retrieval (IR) systems relies on high-quality manual relevance judgments (qrels), which are costly and time-consuming to obtain. While pooling reduces the annotation effort, it results in only partially labeled datasets. Large Language Models (LLMs) offer a promising alternative to reducing reliance on manual judgments, particularly in complex domains like medical case-based retrieval, where relevance assessment requires analyzing both textual and visual information.
    In this work, we explore using a Multimodal Large Language Model (MLLM) to expand relevance judgments, creating a new dataset of automated judgments. Specifically, we employ Gemini 1.5 Pro on the ImageCLEFmed 2013 case-based retrieval task, simulating human assessment through an iteratively refined, structured prompting strategy that integrates binary scoring, instruction-based evaluation, and few-shot learning. We systematically experimented with various prompt configurations to maximize agreement with human judgments.
    To evaluate agreement between the MLLM and human judgments, we use Cohen’s Kappa, achieving a substantial agreement score of 0.6, comparable to inter-annotator agreement typically observed in multimodal retrieval tasks.
    Starting from the original 15,028 manual judgments (4.72\% relevant) across 35 topics, our MLLM-based approach expanded the dataset by over 37x to 558,653 judgments, increasing relevant annotations to 5,950. On average, each medical case query received 15,398 new annotations, with approximately 99\% being non-relevant, reflecting the high sparsity typical in this domain.
%
    Our results demonstrate the potential of MLLMs to scale relevance judgment collection, offering a promising direction for supporting retrieval evaluation in medical and multimodal IR tasks.
\end{abstract}




\maketitle

\vspace{0.5em}
\noindent\textit{Presented at the Third Workshop on Large Language Models for Evaluation in Information Retrieval (LLM4Eval 2025), co-located with the 48th International Conference on Research and Development in Information Retrieval (SIGIR 2025), Padua, Italy, July 17, 2025.}

\section{Introduction}



The evaluation of Information Retrieval (IR) systems is essential for research, as it measures performance and guides the development of improved methods. However, the process of evaluating IR systems comes with challenges, with one of the most significant being the development of relevance judgments (qrels). Under the Cranfield paradigm~\cite{CLEVERDON1967}, IR evaluations rely on test collections comprising a corpus of documents, a set of queries, and relevance judgments, whose quality directly impacts evaluation reliability.

Initially, relevance judgments were assumed to be complete, with all documents reviewed for each query~\cite{CLEVERDON1967}. However, as collections grew in size, this became impractical due to the labor-intensive nature of manual assessments. To mitigate this, pooling~\cite{jones1975report} was introduced, aggregating results from multiple systems to form a subset for judgment~\cite{DBLP:journals/ir/BuckleyDSV07}. While pooling reduces workload, it is influenced by selection biases and system dependency, often resulting in incomplete and potentially unrepresentative judgments~\cite{DBLP:conf/sigir/BuckleyV04}.
These limitations affect IR metrics like MAP and Precision, which either ignore unjudged documents or treat them as non-relevant~\cite{DBLP:conf/sigir/RadlinskiC10}. This can unfairly disadvantage retrieval models and lead to inaccurate evaluations, distorting progress in the field.

Recent advancements in artificial intelligence, particularly in the development of large language models (LLMs), offer a promising avenue for mitigating the limitations of traditional relevance judgment processes~\cite{DBLP:conf/emtcir/AbbasiantaebMAA24}. By harnessing the capabilities of LLMs, the number of relevance judgments available for IR evaluation can be significantly expanded, even from a limited initial sample.
Moreover, multimodal LLMs (MLLMs) extend this potential even further by enabling the assessment of outputs that integrate multiple data types, paving the way for more comprehensive and flexible evaluation frameworks.

This paper explores the application of an MLLM to automatically expand relevance judgments for the ImageCLEFmed 2013 case-based retrieval task dataset~\cite{DBLP:conf/clef/HerreraKDAM13}, where both documents and queries are multimodal, including text and images. 
The proposed approach addresses the challenges of manual relevance judgments by expanding the set of qrels using a dense retrieval method on multimodal data—an approach that remains relatively unexplored for this task. Additionally, this work contributes to the evolution of test collections for evaluating modern retrieval systems, addressing the gap left by existing judgments that fail to capture recent advancements in semantic search. By integrating MLLM-driven expansion methods, we further enhance IR evaluation processes, enabling more comprehensive system assessments while reducing dependence on costly human labor. Moreover, this study advances research on MLLM-as-a-Judge systems by conducting an experiment in this domain, where exploration remains limited.

\Autoref{sec:related-work} provides a review of the literature on the ImageCLEF task and LLM-based evaluators. \Autoref{sec:mllm-judge} describes the approach used to address the small number of qrels in the dataset, detailing the experimentation process that led to the final MLLM-as-a-Judge framework. \Autoref{sec:resources} presents the resources, including a comparison between the original qrels and those expanded using the MLLM-as-a-Judge approach. Finally, \Autoref{sec:conclusions} concludes with key insights and potential directions for future research.
\section{Related Work}\label{sec:related-work}

\subsection{ImageCLEFmed 2013 Case-based Retrieval}\label{sec:related-work:task}


Serving as a long-standing benchmark for medical image retrieval, ImageCLEFmed has driven significant advancements in the field, with annual editions held since 2004~\cite{DBLP:conf/ecir/IonescuMDPIHASSAFIASMFCDSRSBYRDCSPHBKH23}. Early iterations of ImageCLEFmed primarily focused on ad-hoc retrieval tasks, gradually evolving to incorporate case-based retrieval in 2009~\cite{DBLP:journals/cmig/Kalpathy-CramerHDABM15}, which continued until 2013. This shift aimed to better simulate real-world clinical information needs by presenting retrieval tasks as patient cases with associated images and textual descriptions. The evolution of the dataset, from CasImage teaching files~\cite{casImage} to the inclusion of PubMed Central images, introduced new challenges such as filtering non-clinical images and handling compound figures~\cite{DBLP:journals/cmig/Kalpathy-CramerHDABM15}. 

Over its five-year duration, it initially featured only five topics but gradually expanded to 35. The case-based topics underwent relevance assessments by at least two judges to evaluate inter-rater agreement, measured using Cohen's Kappa coefficient. Agreement levels varied across judges and topics, progressively improving over the years, with the average Kappa ranging from 0.14 to 0.67 in the 2011 competition~\cite{DBLP:conf/clef/MullerKEBRBKH09,DBLP:conf/clef/MullerKEBRKH10,DBLP:conf/clef/Kalpathy-CramerMBEHT11}.
Participation also varied, with six to nine groups taking part each year, submitting between 18 and 48 runs. These runs were distributed across three sub-tasks: text, visual, and multimodal retrieval, with textual retrieval consistently receiving the majority of submissions.

A critical aspect of ImageCLEFmed is its rigorous relevance assessment process. It employs clinicians and biomedical informatics students who did not create the topics, thereby reducing bias~\cite{DBLP:books/daglib/0031898}. The use of a ternary relevance scheme (``relevant'', ``partially relevant'', and ``non-relevant'') acknowledges the inherent ambiguity in medical image interpretation~\cite{DBLP:conf/clef/MullerDDKKH07, DBLP:conf/civr/CloughSM04}, which is subsequently converted to binary judgments, treating partially relevant documents as either relevant (“lenient”) or non-relevant (“strict”)~\cite{DBLP:books/daglib/0031898}. However, the judgment process is resource-intensive, requiring significant time and effort from expert judges~\cite{DBLP:conf/amr/MullerCHG06, DBLP:books/daglib/0031898}. The challenge of inter-rater variability is also acknowledged, with studies highlighting the impact of domain expertise on relevance judgments~\cite{DBLP:books/daglib/0031898}. 

This work adopts the established definition of relevance within the ImageCLEFmed framework, particularly for case-based retrieval tasks. Relevance in this context is determined by clinical relevance, judged by clinicians using strict criteria. Since introducing the case-based task, the definition of relevance has been further refined to account for the specific nature of case-based retrieval. This involves providing judges with detailed documentation outlining the criteria for relevance, including examples of what constitutes relevant and non-relevant cases~\cite{DBLP:books/daglib/0031898}. Furthermore, the definition acknowledges the influence of the clinician's expertise on relevance judgments. As highlighted by Müller et al.~\cite{DBLP:books/daglib/0031898}, non-specialists may consider a broader range of cases as relevant, while specialists may require a higher degree of similarity within the same diagnosis group. This nuanced understanding of relevance ensures that the evaluation accurately reflects the information needs of clinicians in a real-world setting.

\subsection{(M)LLM-as-a-Judge}

The evaluation of information retrieval systems has traditionally relied on human assessments to determine the relevance and quality of retrieved documents. While effective, this manual evaluation process is labor-intensive, time-consuming, and often not scalable, particularly impacting large datasets. To address these challenges, researchers have explored the use of Large Language Models as automated judges, coining the term ``LLM-as-a-judge''~\cite{DBLP:conf/nips/ZhengC00WZL0LXZ23}.
LLM-based evaluators~\cite{DBLP:conf/uist/ShankarZHPA24} leverage the language processing capabilities of LLMs to assess documents based on custom criteria specified within an evaluation prompt. This approach has the potential to reduce reliance on extensive human labor and also enables scalable relevance judgments, allowing unjudged documents within collections to be assessed.

In prior work, LLM judges have been categorized based on their input and output format, representing different judgment processes~\cite{DBLP:journals/corr/abs-2411-16594}.
Specifically, there are three types of input formats depending on the number of candidate samples being evaluated. In \textbf{pointwise judgment}, the LLM assesses a single candidate sample in isolation~\cite{DBLP:journals/corr/abs-2304-02554}. In \textbf{pairwise judgment}, two candidate samples are compared to determine relative quality or relevance~\cite{DBLP:conf/nips/ZhengC00WZL0LXZ23}. Lastly, in \textbf{list-wise judgment}, more than two candidates are evaluated together~\cite{DBLP:journals/corr/abs-2412-20061}.

Similarly, the judgment process can be categorized based on the output format, which falls into three main types: score-based, ranking-based, and selection-based judgments. In \textbf{score-based judgment}, each candidate sample is assigned a continuous or discrete score, quantifying its relevance or quality~\cite{DBLP:journals/corr/abs-2407-02694}. In \textbf{ranking-based judgment}, the model outputs an ordered list of candidates according to their relative importance or suitability~\cite{DBLP:journals/tmlr/LiPD24}. Lastly, in \textbf{selection-based judgment}, the LLM identifies one or more optimal candidates from the set, making a direct decision on relevance or preference~\cite{DBLP:conf/nips/YaoYZS00N23}.

With Multimodal Large Language Models (MLLMs) capable of processing and understanding information from multiple modalities such as text, images, and audio, the scope of LLM-as-a-Judge has expanded. Multimodal LLM-as-a-Judge~\cite{DBLP:conf/icml/ChenCZWLZZ00024} systems are designed to assess outputs that combine multiple modalities.
%
This work uses a Multimodal LLM-as-a-Judge approach, employing Gemini 1.5 Pro~\cite{DBLP:journals/corr/abs-2403-05530} for pointwise, score-based evaluation with a binary score system to determine a document's relevance to a given query topic.



\section{MLLM-as-a-Judge for Medical~Case-based~Retrieval}\label{sec:mllm-judge}

%

The ImageCLEFmed 2013 case-based task evaluation was based on a sample of articles obtained through pooling. However, pooled documents were retrieved using sparse methods, as dense or semantic retrieval approaches had not yet been developed. This limits the evaluation of modern retrieval techniques that often surface different sets of potentially relevant documents. Expanding relevance judgments with semantic methods opens the opportunity to make this resource useful for evaluating contemporary retrieval systems.


This study adopts an evaluation-by-criteria approach, inspired by the assessment methodology used in the ImageCLEFmed 2013 case-based retrieval task. Specifically, the LLM is assigned the task of determining whether a given medical article is relevant to a specific medical case (topic), framing the evaluation as a binary classification problem. To this end, we employed Gemini 1.5 Pro~\cite{DBLP:journals/corr/abs-2403-05530} as the MLLM judge, selected for its multimodal capabilities and strong performance, which enable it to process both textual and visual content within the collection. The objective was to approximate the MLLM’s behavior to that of human assessors by providing clear and structured evaluation instructions.

To develop the MLLM judge, we followed an iterative process of designing, testing, and refining evaluation prompts, identifying the versions that yielded the best performance. These prompts were crafted to guide Gemini 1.5 Pro in processing the input and generating relevance judgments. The performance of the MLLM judge was assessed by comparing its outputs against ground truth, with reliability measured through Inter-Annotator Agreement (IAA) between manual judgments and MLLM-generated assessments.

The process of creating the MLLM judge involved the following steps:

\begin{enumerate}
    \item \textbf{Define the evaluation scenario:} Establish the context and objectives of the evaluation task.
    \item \textbf{Prepare the evaluation dataset:} Curate a subset tailored to the requirements of the evaluation.
    \item \textbf{Craft the evaluation prompt:} Design prompts specifying the MLLM’s role and relevance criteria.
    \item \textbf{Evaluate and iterate:} Test the MLLM judge, analyze its performance, and refine the prompts accordingly.
\end{enumerate}

As the case-based retrieval task already included evaluated articles, we constructed a subset comprising up to three relevant and three non-relevant articles for each topic. The choice of three per class reflected a trade-off between achieving representative coverage and maintaining evaluation feasibility. For some topics, fewer assessed articles were available, as detailed in \autoref{tab:qrels}. The final subset included 200 articles across all 35 topics, yielding a total of 202 relevance judgments, as summarized in \autoref{tab:qrels-subset-description}.

\begin{table}
\centering
\caption{\label{tab:qrels-subset-description}MLLM-as-a-Judge evaluation dataset description.}
\begin{tabular}{lr}
\toprule[1pt]
& \textbf{Count} \\
\midrule
\footnotesize{\textit{\textbf{Documents}}} \\ 
\phantom{--}\# articles               & 200     \\
\phantom{--}\# articles images               & 763     \\
\footnotesize{\textit{\textbf{Topics}}} \\ 
\phantom{--}\# cases               & 35     \\
\phantom{--}\# case images               & 76     \\
\footnotesize{\textit{\textbf{Relevance judgments}}} \\ 
\phantom{--}\# annotated qrels & 202 \\
\phantom{--}\# not-relevant & 105 \\
\phantom{--}\# relevant & 97 \\
\phantom{--}Average judgments per query & 5.77 \\
\bottomrule[1pt]
\end{tabular}
\end{table}

We evaluated performance on the constructed test subset by calculating the number and percentage of matching assessments between the task assessors and the MLLM. In addition, we quantified inter-annotator agreement using Cohen’s Kappa coefficient, providing a statistical measure of consistency between MLLM-generated judgments and human annotations.


\subsection{Prompt Engineering}

The effectiveness of (M)LLM judges depends on several critical factors, including model selection, prompt design, and task complexity.
Task complexity, in particular, influences the model’s ability to accurately interpret instructions and directly impacts its alignment with human judgment~\cite{DBLP:journals/corr/abs-2411-15594}. More intricate tasks may pose challenges that reduce output consistency or reliability.

The model’s capabilities—such as its capacity to process information and handle diverse topics—are essential for accurately following and responding to the given instructions. Furthermore, the context window size imposes a practical constraint on the volume of information the model can process at once, thereby limiting the complexity and scope of evaluation prompt design.

Prompt design is pivotal, as variations in structure and phrasing can significantly influence model responses. Well-structured prompts are essential for eliciting accurate and reliable outputs from the model~\cite{DBLP:journals/corr/abs-2411-15594}.

To create the evaluation prompts, we employed several established techniques aimed at optimizing the performance of the MLLM. As LLMs are inherently designed for text generation and are not tuned for consistent or calibrated relevance judgments, we adopted a binary relevance scale (``relevant'' or ``not relevant''). The prompt explicitly defined the criteria for each label and provided guidance on how to handle cases of uncertainty or insufficient information.


To support the reasoning process, we structured the assessment instructions as a sequence of logical steps, guiding the LLM through step-by-step reasoning toward the final decision. This approach helps align the model’s reasoning with that of human evaluators.

Additionally, we experimented with including an example of a medical article relevant to the query topic—sourced from the task’s ground truth relevance judgments (qrels)—by leveraging the few-shot learning technique~\cite{DBLP:conf/nips/BrownMRSKDNSSAA20}. Providing such examples helps the LLM better understand the task context and relevance criteria, thereby improving evaluation accuracy.



\subsection{Judgment Variance by Prompt Changes}



We used Cohen’s Kappa coefficient to measure the agreement between human assessors and the MLLM judge on the constructed test subset. Refinement of the prompts resulted in varying Kappa values, as shown in \autoref{tab:prompts-kappa}, with the specific prompt variants and experimental setups described in the subsequent section. To instruct the MLLM, three key components were required: task instructions, the case presentation, and the article whose relevance to the case was being evaluated. These elements formed the foundation for exploring different prompt configurations.

We analyzed two distinct types of prompts: system prompts and user prompts. System prompts provide a foundational framework that shapes the model’s general behavior. User prompts, on the other hand, consist of task-specific instructions that guide the model’s responses in a more immediate and targeted manner.

Our study compared two approaches: one where the task instructions were provided in a system prompt, followed by user prompts for the case presentation and article, and another where a single user prompt combined the task instructions and case presentation, with subsequent prompts introducing the articles.
The findings revealed that using a system prompt to define the task instructions slightly improved performance compared to setups without a system prompt. This highlights the value of separating foundational guidance into a system prompt to enhance the model’s understanding and task execution.




We further explored the impact of including an example of a relevant article for each case, randomly selected from the ground truth dataset. As noted earlier, the relevance judgments dataset from the ImageCLEFmed task is sparsely populated. Consequently, for 10 out of the 35 topics, the pre-selected example overlapped with the test set, potentially introducing some bias in evaluating the MLLM-as-a-Judge. 
Despite this overlap, our analysis showed that most differences between the zero-shot and few-shot approaches were not in these overlapping articles. Most examples provided in the few-shot approach were already recognized as relevant by the model in the zero-shot setup. However, including these examples in the few-shot approach improved performance considerably, suggesting their role in refining the model's outputs.

Finally, we refined the prompt structure by separating the case presentation prompt from the prompt presenting the relevant article. This adjustment aimed to reduce the complexity of the prompts, ensuring clarity and avoiding the mixing of distinct information, such as the details of the case and associated images. This separation proved effective, yielding improved performance and achieving our experiments' highest Cohen's Kappa value (0.6) with approximately 80\% matching assessments between assessors. The resulting agreement between the task assessors and the MLLM judge reached a substantial level, underscoring the benefits of the different strategies. Notably, our Kappa value surpasses the majority of those observed between human assessors in this task, previously reported in~\Autoref{sec:related-work:task}, and aligns with the highest one achieved.

In the final version, which achieved the highest Kappa score, we employed four distinct prompts detailed in \autoref{app:prompt-templates}: a system prompt providing the MLLM with task instructions (\ref{system-prompt}), a case-presenting prompt (\ref{case-presenting-prompt}), a prompt presenting a relevant article example for the given case (\ref{relevant-example-prompt}), and a prompt introducing the article whose relevance to the case was being evaluated (\ref{article-evaluate-prompt}).

\begin{table}
\centering
\caption{\label{tab:prompts-kappa}Cohen’s Kappa values for different prompt variants with Gemini 1.5 Pro as judge, measuring the level of agreement with the human assessors of the task.}
\begin{tabular}{lc}
\toprule[1pt]
\textbf{Prompt Variant} & \textbf{Kappa}  \\
\midrule
\footnotesize{\textit{\textbf{Zero-shot}}} \\ 
\phantom{--}No system               & 0.5089     \\
\phantom{--}With system             & 0.5198     \\
\footnotesize{\textit{\textbf{Few-shot}}} \\
\phantom{--}Single prompt              & 0.5805      \\
\phantom{--}Separate prompts        & 0.6000      \\
\bottomrule[1pt]
\end{tabular}
\end{table}

\section{Resource Description}\label{sec:resources}

The ImageCLEFmed case-based retrieval task focuses on retrieving medical cases, where each query consists of a multimodal case description that includes both textual and visual information. The retrieval unit is a PubMed article accompanied by its associated images. This task emulates a clinician’s diagnostic workflow by aiming to identify relevant literature from the extensive biomedical content available in PubMed Central.


\subsection{Original ImageCLEFmed 2013 Dataset}

As outlined in \autoref{tab:dataset-description}, the dataset includes article cases, query topics, and relevance judgments. The collection comprises approximately 75,000 articles and 300,000 images from PubMed Central.
The dataset features 35 query topics, each presenting a case description that includes patient demographics, limited symptoms, and test results—such as imaging studies—while omitting the final diagnosis. Due to practical constraints, only a subset of retrieved cases could undergo human review. To address this limitation, a pooling strategy was employed, generating a shortlist of approximately 1,000 cases per topic for detailed evaluation. This process ensured relevance assessments and the establishment of the ground truth for the task, resulting in a total of 15,028 relevance judgments.

\begin{table}
\centering
\caption{\label{tab:dataset-description}ImageCLEFmed 2013 dataset description.}
\begin{tabular}{lr}
\toprule[1pt]
& \textbf{Count} \\
\midrule
\footnotesize{\textit{\textbf{Documents}}} \\ 
\phantom{--}\# articles               & 74,654     \\
\phantom{--}\# article images               & 306,538     \\
\footnotesize{\textit{\textbf{Topics}}} \\ 
\phantom{--}\# cases               & 35     \\
\phantom{--}\# case images               & 76     \\
\footnotesize{\textit{\textbf{Relevance judgments}}} \\ 
\phantom{--}\# annotated qrels & 15,028 \\
\bottomrule[1pt]
\end{tabular}
\end{table}

As shown in \autoref{tab:qrels}, the manual relevance judgments for the ImageCLEFmed 2013 task include 709 documents marked as relevant and 14,319 as not relevant, across 35 query topics. The number of judged articles per topic ranges from 372 to 480. Given that the full collection comprises approximately 75,000 articles, the set of judged documents constitutes a small fraction of the dataset. On average, only about 430 documents—roughly 0.57\% of the collection—were assessed per topic. This highlights a central limitation of the pooling technique: the incompleteness of relevance assessments. Since pooling samples documents based on top-ranked results from submitted systems, it inherently overlooks many unjudged documents, some of which may be relevant. This incompleteness can negatively impact evaluation reliability, as unjudged documents are often implicitly treated as non-relevant by many IR metrics~\cite{DBLP:journals/ires/CloughS13}, potentially introducing bias against systems that retrieve relevant documents outside the original pool.


\begin{table*}
\centering
\caption{\label{tab:qrels}Number of judged articles per query topic in the case-based ImageCLEFmed 2013 task.}
\begin{tabular}{llccccccccc}
\toprule[1pt]
\multicolumn{2}{l}{\textbf{Topic}}                                                 & \textbf{1}  & \textbf{2}  & \textbf{3}  & \textbf{4}  & \textbf{5}  & \textbf{6}  & \textbf{7}  & \textbf{8}  & \textbf{9}     \\
\multicolumn{1}{c}{\multirow{2}{*}{\textbf{Articles}}} & Relevant     & 21          & 3           & 3           & 4           & 34          & 54          & 33          & 40          & 3              \\ \cline{2-2}
\multicolumn{1}{c}{}                                              & Not relevant & 369         & 477         & 468         & 476         & 377         & 377         & 410         & 364         & 393            \\ \midrule
\multicolumn{2}{l}{\textbf{Topic}}                                                 & \textbf{10} & \textbf{11} & \textbf{12} & \textbf{13} & \textbf{14} & \textbf{15} & \textbf{16} & \textbf{17} & \textbf{18}    \\
\multirow{2}{*}{\textbf{Articles}}                     & Relevant     & 1           & 1           & 3           & 24          & 58          & 5           & 2           & 1           & 10             \\ \cline{2-2}
                                                                  & Not relevant & 415         & 433         & 380         & 392         & 398         & 461         & 440         & 458         & 381            \\ \midrule
\multicolumn{2}{l}{\textbf{Topic}}                                                 & \textbf{19} & \textbf{20} & \textbf{21} & \textbf{22} & \textbf{23} & \textbf{24} & \textbf{25} & \textbf{26} & \textbf{27}    \\
\multirow{2}{*}{\textbf{Articles}}                     & Relevant     & 17          & 32          & 32          & 53          & 38          & 11          & 3           & 101         & 8              \\ \cline{2-2}
                                                                  & Not relevant & 410         & 378         & 439         & 406         & 359         & 452         & 465         & 333         & 397            \\ \midrule
\multicolumn{2}{l}{\textbf{Topic}}                                                 & \textbf{28} & \textbf{29} & \textbf{30} & \textbf{31} & \textbf{32} & \textbf{33} & \textbf{34} & \textbf{35} & {\textbf{Total}} \\
\multirow{2}{*}{\textbf{Articles}}                     & Relevant     & 7           & 15          & 41          & 2           & 26          & 4           & 9           & 10          & 709            \\ \cline{2-2}
                                                                  & Not relevant & 371         & 433         & 401         & 448         & 407         & 368         & 410         & 373         & 14,319         \\ \bottomrule[1pt]
\end{tabular}
\end{table*}


\subsection{Extended Relevance Judgments}


As mentioned in \Autoref{sec:mllm-judge}, our goal was to evaluate multimodal dense retrieval approaches for this task. However, the limited size of the ground truth qrels dataset constrained our evaluation. The original ImageCLEFmed 2013 dataset contained 15,028 relevance judgments, with only 709 (4.72\%) marked as relevant. To address this limitation, we chose to judge the missing articles from our retrieval set of experiments, significantly expanding the relevance judgments using the MLLM-as-a-Judge method, as shown in \autoref{tab:distribution-qrels}.

Following this expansion, the qrels dataset grew over 37 times in size, reaching 558,653 relevance judgments.
In some cases, Google's generative AI model flagged prompts as containing prohibited content, preventing the model from assessing relevance. To handle this, we applied a conservative approach, marking these articles as not relevant. Due to these safety filters, 1,195 judgments were classified as not relevant. Although the percentage of relevant judgments decreased to 1.07\%, the total count increased more than eightfold, reaching 5,950. This reflects a common trend in relevance assessments, where the majority of annotated qrels are typically classified as not relevant~\cite{DBLP:conf/sigir/ChenLHL0SEZK024}.

\begin{table}
\caption{\label{tab:distribution-qrels}Distribution of Qrels.}
\centering
\resizebox{\columnwidth}{!}{%
\begin{tabular}{lrr}
\toprule[1pt]
\textbf{Qrel} & \textbf{ImageCLEFmed 2013} & \textbf{MLLM-as-a-Judge} \\
\midrule
Total & 15,028\phantom{ (95.28\%)} & 558,653\phantom{ (98.55\%)} \\
\phantom{11:}0: not-relevant & 14,319 (95.28\%) & 552,703 (98.93\%) \\
\phantom{11:}1: relevant & 709 (\phantom{0}4.72\%) & 5,950 (\phantom{0}1.07\%) \\
\bottomrule[1pt]
\end{tabular}}
\end{table}

\begin{figure}
\centering
\includegraphics[width=0.475\textwidth]{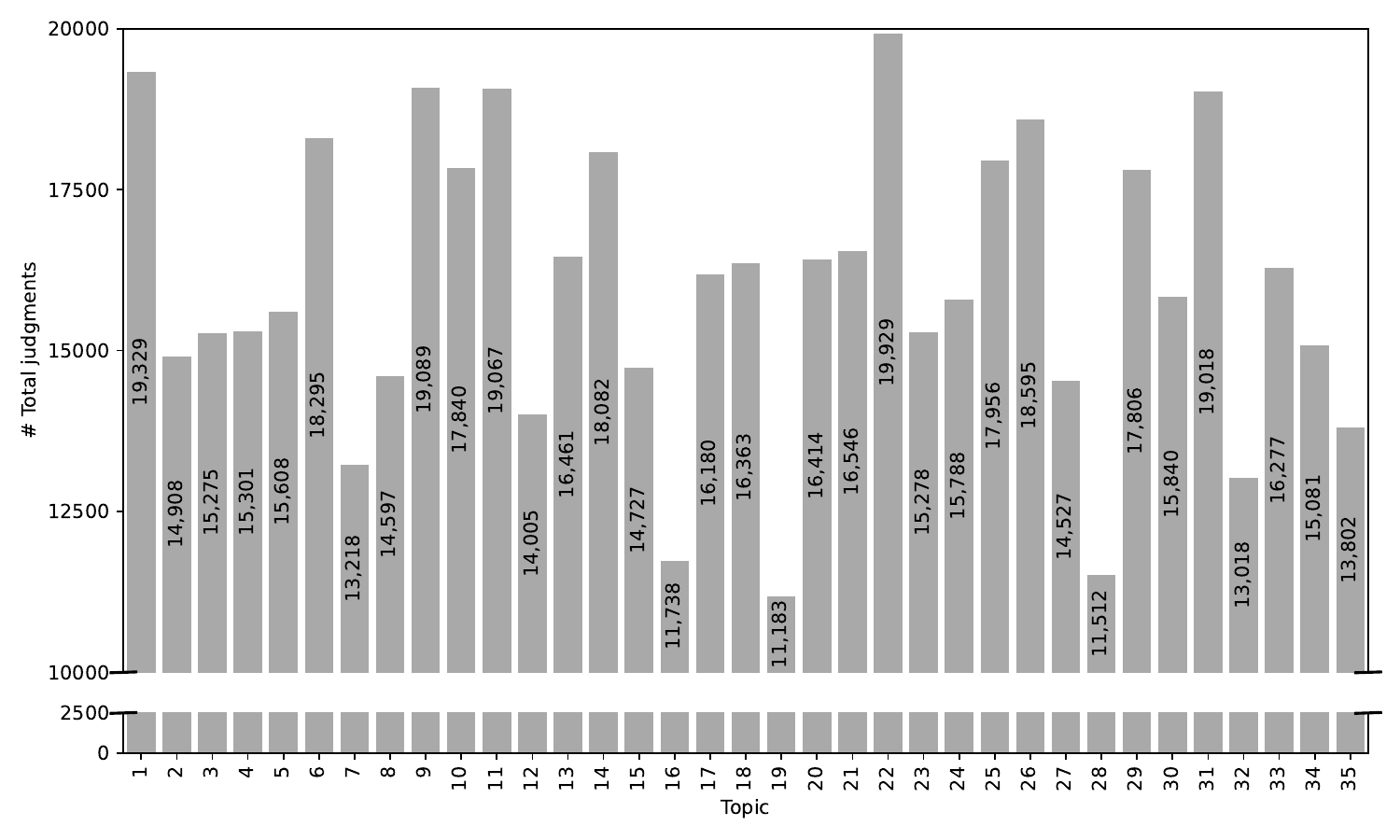}
\caption{\label{fig:mllm-total-judged}Total number of judgments per topic of after the MLLM-as-a-Judge qrels expansion.}
\Description{A barplot of the number of relevance judgments per topic after the MLLM-as-a-Judge qrels expansion.}
\end{figure}

\begin{figure}
\centering
\includegraphics[width=0.475\textwidth]{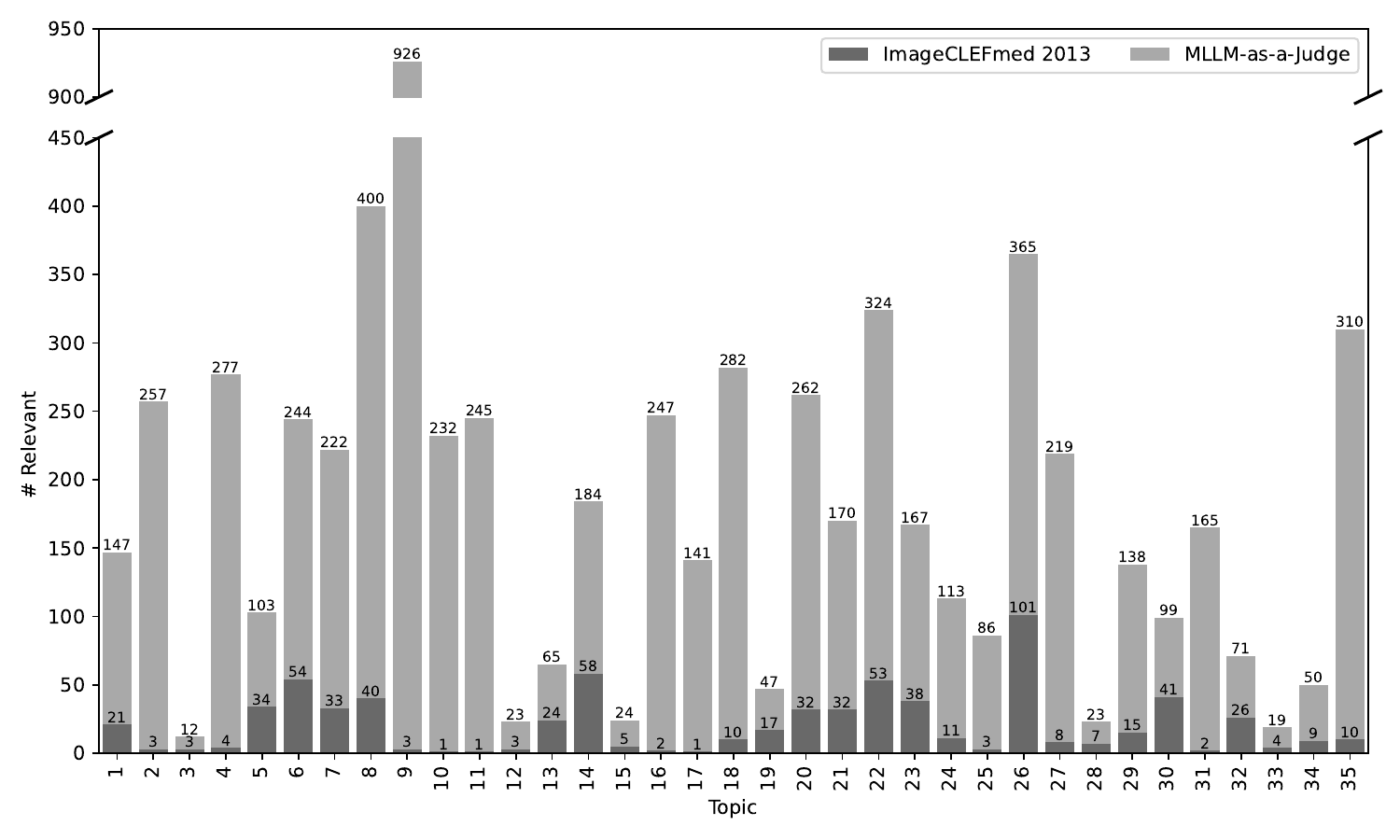}
\caption{\label{fig:relevant-diff}Difference between the number of relevant judgments between the original ImageCLEFmed 2013 qrels and the MLLM-as-a-Judge qrels expansion.}
\Description{A stacked barplot of the difference between the number of relevant articles between the original ImageCLEFmed 2013 qrels and the MLLM-as-a-Judge qrels expansion.}
\end{figure}

\autoref{fig:mllm-total-judged} shows the total number of relevance judgments after the MLLM-as-a-Judge qrels expansion. On average, each topic received 15,398 annotations, with 108 marked as relevant and 15,381 as not relevant.
As illustrated in \autoref{fig:relevant-diff}, the number of relevant judgments increased by a factor of six on average compared to the original qrels, with per-topic gains ranging from less than twofold to over 300-fold. For instance, topic 30 saw a modest increase from 41 to 99 relevant documents (2.4×), while topic 9 exhibited the largest increase, from 3 to 926.
Overall, the expansion yielded an eightfold increase in the number of relevant judgments and a 39-fold increase in non-relevant ones. As a result, the proportion of the collection judged per topic rose from approximately 0.57\% to 21\% on average.

\subsection{Resource Access}\label{sec:availability}

%

The additional relevance judgments (qrels) for the ImageCLEFmed 2013 case-based retrieval task, along with a detailed resource description and the source code used to generate the judgments, are publicly available in an online repository\footnote{\url{https://github.com/catarinaopires/mllm-as-a-judge}} and archived with a persistent DOI for citation purposes~\cite{mllm-expansion_qrel_2013_case_based}.

\section{Conclusion and Future Work}\label{sec:conclusions}

This study demonstrated the potential of using a Multimodal Large Language Model (MLLM) to expand relevance judgments for the ImageCLEFmed 2013 case-based retrieval task. Using an MLLM-as-a-Judge approach, we scaled the initial dataset from 15,028 to 558,653 judgments, increasing the average proportion of the collection judged per query topic from approximately 0.57\% to 21\%.
This expansion provides a more extensive ground truth for evaluating IR systems in the medical domain.
The findings underscore the potential of LLMs, particularly multimodal ones, in reducing the high costs and time constraints associated with manual qrel annotation.

Prompt engineering played a crucial role, as different prompts yielded varying agreement levels compared to manual judgments. While our best setup achieved a substantial agreement score of 0.6, further refinements may yield even higher alignment with expert judgments. For instance, having access to the original task guidelines—used to instruct human assessors on evaluating article relevance to a case—could enhance our approach.

A key limitation of using LLMs to generate relevance judgments stems from the fact that the relevance labels are produced by the model itself, meaning any evaluation is inherently constrained by the capabilities and biases of that model. This limits what can be measured to the level of the generating model~\cite{Soboroff_2025}. However, as we noted, including relevant documents in the prompts can enhance the LLM's performance as a judge.
While LLM-generated relevance assessments have limitations due to the inherent uncertainty of the model’s generative nature, which causes them to fall short of the quality of expert-curated judgments, they still offer advantages over human labor. They can be comparable to crowdsourced assessments and serve as valuable noisy training data~\cite{DBLP:conf/sigir/0001SC024}.

Our results indicate that while the majority of generated annotations remain non-relevant, this is consistent with real-world IR tasks, where relevant documents constitute a small fraction of the total. The large-scale expansion of qrels facilitated by our method can significantly enhance IR evaluation, allowing for more robust benchmarking of retrieval systems.

Furthermore, this research contributes to the growing exploration of MLLM-as-a-Judge methodologies, with a particular focus on medical IR—an area where automated relevance assessment remains underexplored. By applying multimodal LLMs to case-based retrieval, our work contributes to advancing evaluation frameworks capable of handling both textual and visual modalities.
Future work should focus on refining prompt strategies, as they have proven to be a critical and decisive aspect of the process. Additionally, experimenting with different MLLMs, including those fine-tuned to the medical domain and equipped with expert knowledge, could further enhance the model's performance and domain-specific relevance.


In conclusion, this study highlights the scalability and practical value of using MLLMs to expand qrels for relevance assessment in information retrieval. Leveraging advanced LLMs enables the enhancement of evaluation frameworks and broader coverage—particularly in domains like medical retrieval, where manual annotation is both costly and limited.

\begin{acks}
This work was financed by Component 5 - Capitalization and Business Innovation, integrated into the Resilience Dimension of the Recovery and Resilience Plan, within the scope of the Recovery and Resilience Mechanism (MRR) of the European Union (EU), framed under Next Generation EU, for the period 2021–2026, within the project HfPT (reference 41).
Sérgio Nunes acknowledges the use of computational resources provided by the project ``PTicola – Increasing Computationally Language Resources for Portuguese'' (reference \url{https://doi.org/10.54499/CPCA-IAC/AV/594794/2023}).
\end{acks}

\bibliographystyle{ACM-Reference-Format}
\bibliography{pires2025-mllm-qrel-expansion}

\appendix
\onecolumn
\section{Prompt Templates}\label{app:prompt-templates}

\begin{promptbox}[System prompt of MLLM in judging.]{system-prompt}
You are an expert judge evaluating the relevance of medical articles for a case-based retrieval task. Your role is to determine if a given article is relevant to a patient's case.

\textbf{Case Information:}

- Textual Description: A narrative detailing the patient's demographics and symptoms.

- Images: Imaging studies (e.g., CT scans, MRIs) associated with the case.

\textbf{Article Information:}

- Title: Title of the medical article.

- Authors: Names of the contributing authors.

- Fulltext: The complete text of the medical article.

- Images: Images included in the article.

- Image captions: Captions describing the article images.

\phantom{1}

\textbf{Evaluation Criteria:}

Assess the article based on these criteria:

- Differential Diagnosis: Does the article provide information that could help in identifying or differentiating the possible diagnoses for the patient's condition?

- Clinical Evidence: Does the article discuss similar cases, findings, or treatments that align with the patient's case description and images?

- Specificity and Accuracy: Does the article contain detailed, accurate, and medically relevant content directly applicable to the case?

\phantom{1}

\textbf{Relevance Score:}

Assign a binary score to the article:

~1: Relevant: The article provides information that could be useful for understanding the patient's condition, making a diagnosis, or planning their treatment.

~0: Not Relevant: The article does not provide useful information related to the patient's case.

\phantom{1}

\textbf{Instructions:}

1. Carefully read the textual description of the patient case and examine the associated images (if available).

2. Review the provided example of a relevant article to understand the standard for relevance.

3. Review the content of the retrieved article.

4. Assign a relevance score (1 or 0) based on the criteria above.

5. If you are uncertain about the article's relevance, err on the side of caution and assign a score of 0.

\phantom{1}

Focus on providing objective, evidence-based judgments that would genuinely assist a clinician in diagnosing and managing this patient.

Wait for the presentation of the patient’s case and an example of a relevant article before proceeding with your evaluation.
\end{promptbox}

\begin{promptbox}[Case presenting prompt of MLLM in judging.]{case-presenting-prompt}
The patient's case to perform the judgments against is the following:

\phantom{1}

Case Textual description: \{case\_description\}

Case Images: \{case\_images\}

\phantom{1}

Wait for an example of a relevant article before proceeding with your evaluation.
\end{promptbox}

\clearpage

\begin{promptbox}[Relevant article example presenting prompt of MLLM in judging.]{relevant-example-prompt}
To guide your evaluation, here is an example of a medical article that is considered relevant to this patient’s case:

\phantom{1}

Medical article title: \{article\_title\}

Medical article authors: \{article\_authors\}

Medical article abstract: \{article\_abstract\}

Medical article full text: \{article\_fulltext\}

Medical article images: \{article\_images\}

Medical article image captions: \{article\_captions\}

\phantom{1}

Assess whether the following articles are relevant to this patient's case. 
Only answer 0 or 1.
\end{promptbox}

\begin{promptbox}[Article evaluation prompt of MLLM in judging.]{article-evaluate-prompt}
Based on the instructions given, determine if the following article is relevant to the given patient's case. Only answer 0 or 1.

\phantom{1}

Medical article title: \{article\_title\}

Medical article authors: \{article\_authors\}

Medical article abstract: \{article\_abstract\}

Medical article full text: \{article\_fulltext\}

Medical article images: \{article\_images\}

Medical article image captions: \{article\_captions\}
\end{promptbox}

\end{document}